\documentclass{article}%
\usepackage{amsmath}
\usepackage{amsfonts}
\usepackage{amssymb}
\usepackage{graphicx}%
\newtheorem{theorem}{Theorem}

\newtheorem{definition}[theorem]{Definition}

\newtheorem{lemma}[theorem]{Lemma}

\newtheorem{proposition}[theorem]{Proposition}

\newenvironment{proof}[1][Proof]{\noindent\textbf{#1.} }{\ \rule{0.5em}{0.5em}}
\begin{document}

\title{Quantum Lower Bound for Recursive Fourier Sampling}
\author{Scott Aaronson\\Institute for Advanced Study, Princeton\\aaronson@ias.edu}
\date{}
\maketitle

\begin{abstract}
One of the earliest quantum algorithms was discovered by Bernstein and
Vazirani, for a problem called Recursive Fourier Sampling. \ This paper shows
that the Bernstein-Vazirani algorithm is not far from optimal. \ The moral is
that the need to \textquotedblleft uncompute\textquotedblright\ garbage can
impose a fundamental limit on efficient quantum computation. \ The proof
introduces a new parameter of Boolean functions called the \textquotedblleft
nonparity coefficient,\textquotedblright\ which might be of independent interest.

\end{abstract}

Like a classical algorithm, a quantum algorithm can solve problems
recursively by calling itself as a subroutine. \ When this is done,
though, the algorithm typically needs to call itself \textit{twice}
for each subproblem to be solved. \ The second call's purpose is to
uncompute `garbage' left over by the first call, and thereby enable
interference between different branches of the computation. \ Of
course, a factor of $2$ increase in running time hardly seems like a
big deal, when set against the speedups promised by quantum
computing. \ The problem is that these factors of $2$ multiply, with
each level of recursion producing an additional factor. \ Thus, one
might wonder whether the uncomputing\ step is really necessary, or
whether a cleverly designed algorithm might avoid it. \ This paper
gives the first nontrivial example in which recursive uncomputation
is provably necessary.

The example concerns a long-neglected problem called \textit{Recursive Fourier
Sampling} (henceforth $\operatorname*{RFS}$),\ which was introduced by
Bernstein and Vazirani \cite{bv} in 1993 to prove the first oracle separation
between $\mathsf{BPP}$ and \thinspace$\mathsf{BQP}$. \ Many surveys on quantum
computing pass directly from the Deutsch-Jozsa algorithm \cite{dj}\ to the
dramatic results of Simon \cite{simon}\ and Shor \cite{shor}, without even
mentioning $\operatorname*{RFS}$. \ There are two likely reasons for this
neglect. \ First, the $\operatorname*{RFS}$\ problem seems artificial. \ It
was introduced for the sole purpose of proving an oracle result, and is unlike
all other problems for which a quantum speedup is known. \ (I will define
$\operatorname*{RFS}$ in Section \ref{PRELIMRFS}; but for now, it involves a
tree of depth $\log n$, where each vertex is labeled with a function to be
evaluated via a Fourier transform.) \ Second, the speedup for
$\operatorname*{RFS}$\ is only quasipolynomial ($n$ versus $n^{\log n}$),
rather than exponential as for the period-finding and hidden subgroup problems.

Nevertheless, I believe that $\operatorname*{RFS}$ merits renewed
attention---for it serves as an important link between quantum
computing and the ideas of classical complexity theory. \ One reason
is that, although other problems in $\mathsf{BQP}$---such as the
factoring, discrete logarithm, and `shifted Legendre symbol'
problems \cite{dhi}---are thought to be classically intractable,\
these problems are quite low-level by complexity-theoretic
standards. \ They, or their associated decision problems, are in
$\mathsf{NP}\cap\mathsf{coNP}$.\footnote{For the shifted Legendre
symbol problem, this is true assuming a number-theoretic conjecture
of Boneh
and Lipton \cite{bl}.} \ By contrast, Bernstein and Vazirani \cite{bv}%
\ showed that, as an oracle problem, $\operatorname*{RFS}$\ lies
outside $\mathsf{NP}$ and even $\mathsf{MA}$ (the latter result is
unpublished, though not difficult). \ Subsequently Watrous
\cite{watrous}\ gave an oracle $A$, based on an unrelated problem,
for which $\mathsf{BQP}^{A}\not \subset
\mathsf{MA}^{A}$.\footnote{Actually, to place $\mathsf{BQP}$ outside
$\mathsf{MA}$ relative to an oracle, it suffices to consider the
complement of Simon's problem (\textquotedblleft Does $f\left(
x\right) =f\left(  x\oplus s\right)  $ only when
$s=0$?\textquotedblright).} \ Also,
Green and Pruim \cite{gp}\ gave an oracle $B$\ for which $\mathsf{BQP}%
^{B}\not \subset \mathsf{P}^{\mathsf{NP}^{B}}$. \ However, Watrous'
problem was shown by Babai \cite{babai:am}\ to be in $\mathsf{AM}$,
while Green and Pruim's problem is in $\mathsf{BPP}$. \ Thus,
neither problem can be used to place $\mathsf{BQP}$\ outside higher
levels of the polynomial hierarchy $\mathsf{PH}$.

On the other hand, Umesh Vazirani and others have conjectured that
$\operatorname*{RFS}$\ is not in $\mathsf{PH}$, from which it would follow
that there exists an oracle $A$ relative to which $\mathsf{BQP}^{A}%
\not \subset \mathsf{PH}^{A}$. \ Proving this is, in my view, one of the
central open problems in quantum complexity theory. \ Its solution seems
likely to require novel techniques for constant-depth circuit lower
bounds.\footnote{For the $\operatorname*{RFS}$\ function can be represented by
a low-degree real polynomial---this follows from the existence of a
polynomial-time quantum algorithm for $\operatorname*{RFS}$, together with the
result of Beals et al.\ \cite{bbcmw}\ relating quantum algorithms to
low-degree polynomials. \ As a result, the circuit lower bound technique of
Razborov \cite{razborov:ac0}\ and Smolensky \cite{smolensky}, which is based
on the nonexistence of low-degree polynomials, seems unlikely to work. \ Even
the random restriction method of Furst et al.\ \cite{fss}\ can be related to
low-degree polynomials, as shown by Linial et al.\ \cite{lmn}.}

In this paper I examine the\ $\operatorname*{RFS}$\ problem from a
different angle. \ Could Bernstein and Vazirani's quantum algorithm
for $\operatorname*{RFS}$\ be improved even further, to give an
\textit{exponential} speedup over the classical algorithm? \ And
could we use $\operatorname*{RFS}$, not merely to place
$\mathsf{BQP}$\ outside of
$\mathsf{PH}$ relative to an oracle, but to place it outside of $\mathsf{PH}%
$\ with (say) a logarithmic number of alternations?

My answer to both questions is a strong `no.' \ I study a large class of
variations on $\operatorname*{RFS}$, and show that all of them fall into one
of two classes:

\begin{enumerate}
\item[(1)] a trivial class, for which there exists a classical algorithm
making only one query, or

\item[(2)] a nontrivial class, for which any quantum algorithm needs
$2^{\Omega\left(  h\right)  }$\ queries, where $h$ is the height of the tree
to be evaluated. \ (By comparison, the Bernstein-Vazirani algorithm uses
$2^{h}$ queries, because of its need to uncompute garbage recursively at each
level of the tree.)
\end{enumerate}

Since $n^{h}$\ queries always suffice classically, this dichotomy theorem
implies that the speedup afforded by quantum computers is at most
quasipolynomial. \ It also implies that (nontrivial) $\operatorname*{RFS}$\ is
solvable in quantum polynomial time only when $h=O\left(  \log n\right)  $.

The plan is as follows. \ In Section \ref{PRELIMRFS}, I define the
$\operatorname*{RFS}$ problem, and give Bernstein and\ Vazirani's
quantum algorithm for solving it. \ In Section \ref{LOWERRFS}, I use
the adversary method of Ambainis \cite{ambainis}\ to prove a lower
bound on the quantum query complexity of any $\operatorname*{RFS}$
variant. \ This bound, however, requires a parameter that I call the
\textquotedblleft nonparity coefficient\textquotedblright\ to be
large. \ Intuitively, given a Boolean function $g:\left\{
0,1\right\}  ^{n}\rightarrow\left\{  0,1\right\}  $, the nonparity
coefficient measures how far $g$ is from being the parity of some
subset of its input bits---not under the uniform distribution over
inputs (the standard assumption in Fourier analysis), but under an
adversarial distribution. \ The crux of the argument is that
\textit{either} the nonparity coefficient is zero (meaning the
$\operatorname*{RFS}$\ variant in question is trivial), or else it
is bounded below by a positive constant. \ This statement is proved
in Section \ref{LOWERRFS}, and seems like it might be of independent
interest. \ Section \ref{OPENRFS} concludes with some open problems.

\section{Preliminaries\label{PRELIMRFS}}

In ordinary Fourier sampling, we are given oracle access to a Boolean function
$A:\left\{  0,1\right\}  ^{n}\rightarrow\left\{  0,1\right\}  $,\ and are
promised that there exists a secret string $s\in\left\{  0,1\right\}  ^{n}%
$\ such that $A\left(  x\right)  =s\cdot x\left(  \operatorname{mod}2\right)
$\ for all $x$. \ The problem is to find $s$---or rather, since we need a
problem with Boolean output, the problem is to return $g\left(  s\right)  $,
where $g:\left\{  0,1\right\}  ^{n}\rightarrow\left\{  0,1\right\}  $\ is some
known Boolean function. \ We can think of $g\left(  s\right)  $\ as the
\textquotedblleft hard-core bit\textquotedblright\ of $s$, and can assume that
$g$ itself is efficiently computable, or else that we are given access to an
oracle for $g$.

To obtain a height-$2$ recursive Fourier sampling tree, we simply compose this
problem. \ That is, we are no longer given direct access to $A\left(
x\right)  $, but instead are promised that $A\left(  x\right)  =g\left(
s_{x}\right)  $, where $s_{x}\in\left\{  0,1\right\}  ^{n}$ is the secret
string for another Fourier sampling problem. \ A query then takes the form
$\left(  x,y\right)  $, and produces as output $A_{x}\left(  y\right)
=s_{x}\cdot y\left(  \operatorname{mod}2\right)  $. \ As before, we are
promised that there exists an $s$ such that $A\left(  x\right)  =s\cdot
x\left(  \operatorname{mod}2\right)  $ for all $x$, meaning that the $s_{x}%
$\ strings must be chosen consistent with this promise. \ Again we must return
$g\left(  s\right)  $.

Continuing, we can define height-$h$ recursive Fourier sampling, or
$\operatorname*{RFS}_{h}$, recursively as follows. \ We are given oracle
access to a function $A\left(  x_{1},\ldots,x_{h}\right)  $ for all
$x_{1},\ldots,x_{h}\in\left\{  0,1\right\}  ^{n}$,\ and are promised that

\begin{enumerate}
\item[(1)] for each fixed $x_{1}^{\ast}$, $A\left(  x_{1}^{\ast},x_{2}%
,\ldots,x_{h}\right)  $\ is an instance of $\operatorname*{RFS}_{h-1}$ on
$x_{2},\ldots,x_{h}$, having answer bit $b\left(  x_{1}^{\ast}\right)
\in\left\{  0,1\right\}  $; and

\item[(2)] there exists a secret string $s\in\left\{  0,1\right\}  ^{n}$\ such
that $b\left(  x_{1}^{\ast}\right)  =s\cdot x_{1}^{\ast}\left(
\operatorname{mod}2\right)  $\ for each $x_{1}^{\ast}$.
\end{enumerate}

Again the answer bit to be returned is $g\left(  s\right)  $. \ Note
that $g$ is assumed to be the same everywhere in the tree---though
using the techniques in this paper, it would be straightforward to
generalize to the case of different $g$'s. \ As an example that will
be used later, we could take $g\left(  s\right)
=g_{\operatorname{mod}3}\left(  s\right)  $, where
$g_{\operatorname{mod}3}\left(  s\right)  =0$ if $\left\vert
s\right\vert
\equiv0\left(  \operatorname{mod}3\right)  $ and $g_{\operatorname{mod}%
3}\left(  s\right)  =1$ otherwise, and $\left\vert s\right\vert $\ denotes the
Hamming weight of $s$. \ We do not want to take $g$ to be the parity of $s$,
for if we did then $g\left(  s\right)  $\ could be evaluated using a single
query. \ To see this, observe that if $x$\ is the all-$1$'s string, then
$s\cdot x\left(  \operatorname{mod}2\right)  $\ is the parity of $s$.

By an `input,' I will mean a complete assignment for the $\operatorname*{RFS}$
oracle (that is, $A\left(  x_{1},\ldots,x_{h}\right)  $\ for all $x_{1}%
,\ldots,x_{h}$). \ I will sometimes refer also to an `$\operatorname*{RFS}$
tree,' where each vertex at distance $\ell$ from the root has a label
$x_{1},\ldots,x_{\ell}$. \ If $\ell=h$\ then the vertex is a leaf; otherwise
it has $2^{n}$\ children, each with a label $x_{1},\ldots,x_{\ell},x_{\ell+1}$
for some $x_{\ell+1}$. \ The subtrees of the tree just correspond to the
sub-instances of $\operatorname*{RFS}$.

Bernstein and Vazirani \cite{bv}\ showed that $\operatorname*{RFS}_{\log n}$,
or $\operatorname*{RFS}$ with height $\log n$ (all logarithms are base $2$),
is solvable on a quantum computer in time polynomial in $n$. \ I include a
proof for completeness. \ Let $A=\left(  A_{n}\right)  _{n\geq0}$ be an oracle
that, for each $n$, encodes an instance of $\operatorname*{RFS}_{\log n}%
$\ whose answer is $\Psi_{n}$. \ Then let $L_{A}$ be the unary language
$\left\{  0^{n}:\Psi_{n}=1\right\}  $.

\begin{lemma}
\label{alg}$L_{A}\in\mathsf{EQP}^{A}\subseteq\mathsf{BQP}^{A}$ for any choice
of $A.$
\end{lemma}

\begin{proof}
$\operatorname*{RFS}_{1}$ can be solved exactly in four queries, with no
garbage bits left over. \ The algorithm is as follows: first prepare the state%
\[
2^{-n/2}\sum_{x\in\left\{  0,1\right\}  ^{n}}\left\vert x\right\rangle
\left\vert A\left(  x\right)  \right\rangle ,
\]
using one query to $A$. \ Then apply a phase flip conditioned on $A\left(
x\right)  =1$, and uncompute $A\left(  x\right)  $\ using a second query,
obtaining%
\[
2^{-n/2}\sum_{x\in\left\{  0,1\right\}  ^{n}}\left(  -1\right)  ^{A\left(
x\right)  }\left\vert x\right\rangle .
\]
Then apply a Hadamard gate to each bit of the $\left\vert x\right\rangle
$\ register. \ It can be checked that the resulting state is simply
$\left\vert s\right\rangle $. \ One can then compute $\left\vert
s\right\rangle \left\vert g\left(  s\right)  \right\rangle $ and uncompute
$\left\vert s\right\rangle $\ using two more queries to $A$, to obtain
$\left\vert g\left(  s\right)  \right\rangle $. \ To solve $RFS_{\log
n}\left(  n\right)  $, we simply apply the above algorithm recursively at each
level of the tree. \ The total number of queries used is $4^{\log n}=n^{2}$.

One can further reduce the number of queries to $2^{\log n}=n$\ by using the
\textquotedblleft one-call kickback trick,\textquotedblright\ described by
Cleve et al.\ \cite{cemm}. \ Here one prepares the state%
\[
2^{-n/2}\sum_{x\in\left\{  0,1\right\}  ^{n}}\left\vert x\right\rangle
\otimes\frac{\left\vert 1\right\rangle -\left\vert 0\right\rangle }{\sqrt{2}}%
\]
and then exclusive-$OR$'s $A\left(  x\right)  $\ into the second register.
\ This induces the desired phase $\left(  -1\right)  ^{A\left(  x\right)  }%
$\ without the need to uncompute $A\left(  x\right)  $. \ However, one still
needs to uncompute $\left\vert s\right\rangle $\ after computing $\left\vert
g\left(  s\right)  \right\rangle $.
\end{proof}

A remark on notation: to avoid confusion with subscripts, I denote the
$i^{th}$\ bit of string $x$ by $x\left[  i\right]  $.

\section{Quantum Lower Bound\label{LOWERRFS}}

In this section I prove a lower bound on the quantum query complexity of
$\operatorname*{RFS}$. \ Crucially, the bound should hold for any nontrivial
one-bit function of the secret strings, not just a specific function such as
$g_{\operatorname{mod}3}\left(  s\right)  $ defined in Section \ref{PRELIMRFS}%
. \ Let $\operatorname*{RFS}_{h}^{g}$\ be height-$h$ recursive Fourier
sampling in which the problem at each vertex is to return $g\left(  s\right)
$. \ The following notion turns out to be essential.

\begin{definition}
Given a Boolean function $g:\left\{  0,1\right\}  ^{n}\rightarrow\left\{
0,1\right\}  $ (partial or total), the \textit{nonparity coefficient}
$\mu\left(  g\right)  $ is the largest $\mu^{\ast}$\ for which\ there exist
distributions $D_{0}$\ over the $0$-inputs of $g$, and $D_{1}$\ over the
$1$-inputs, such that for all $z\in\left\{  0,1\right\}  ^{n}$, all $0$-inputs
$\widehat{s}_{0}$, and all $1$-inputs $\widehat{s}_{1}$, we have%
\[
\Pr_{s_{0}\in D_{0},s_{1}\in D_{1}}\left[  s_{0}\cdot z\equiv\widehat{s}%
_{1}\cdot z\left(  \operatorname{mod}2\right)  \,\,\,\vee\,\,\,s_{1}\cdot
z\equiv\widehat{s}_{0}\cdot z\left(  \operatorname{mod}2\right)  \right]
\geq\mu^{\ast}\text{.}%
\]

\end{definition}

Loosely speaking, the nonparity coefficient is high if there exist
distributions over $0$-inputs and $1$-inputs that make $g$ far from being a
parity function of a subset of input bits. \ The following proposition
develops some intuition about $\mu\left(  g\right)  $.

\begin{proposition}
\quad

\begin{enumerate}
\item[(i)] $\mu\left(  g\right)  \leq3/4$ for all nonconstant $g$.

\item[(ii)] $\mu\left(  g\right)  =0$\ if and only if $g$ can be written as
the parity (or the NOT of the parity) of a subset $B$ of input bits.
\end{enumerate}
\end{proposition}

\begin{proof}
\quad

\begin{enumerate}
\item[(i)] Given any $s_{0}\neq\widehat{s}_{1}$ and $s_{1}\neq\widehat{s}_{0}%
$, a uniform random $z$ will satisfy%
\[
\Pr_{z}\left[  s_{0}\cdot z\not \equiv \widehat{s}_{1}\cdot z\left(
\operatorname{mod}2\right)  \,\,\,\wedge\,\,\,s_{1}\cdot z\not \equiv
\widehat{s}_{0}\cdot z\left(  \operatorname{mod}2\right)  \right]  \geq
\frac{1}{4}\text{.}%
\]
(If $s_{0}\oplus\widehat{s}_{1}=s_{1}\oplus\widehat{s}_{0}$\ then this
probability will be $1/2$; otherwise it will be $1/4$.) \ So certainly there
is a fixed choice of $z$\ that works for random $s_{0}$\ and $s_{1}$.

\item[(ii)] For the `if' direction, take $z\left[  i\right]  =1$ if and only
if $i\in B$, and choose $\widehat{s}_{0}$\ and $\widehat{s}_{1}$\ arbitrarily.
\ This ensures that $\mu^{\ast}=0$. \ For the `only if' direction, if
$\mu\left(  g\right)  =0$, we can choose $D_{0}$ to have support on all
$0$-inputs, and $D_{1}$\ to have support on all $1$-inputs. \ Then there must
be a $z$\ such that $s_{0}\cdot z$\ is constant as we range over $0$-inputs,
and $s_{1}\cdot z$\ is constant as we range over $1$-inputs. \ Take $i\in
B$\ if and only if $z\left[  i\right]  =1$.
\end{enumerate}
\end{proof}

If $\mu\left(  g\right)  =0$, then $\operatorname*{RFS}_{h}^{g}$\ is easily
solvable using a single classical query. \ Theorem \ref{rfslb} will show that
for all $g$ (partial or total),%
\[
\operatorname*{Q}\nolimits_{2}\left(  \operatorname*{RFS}\nolimits_{h}%
^{g}\right)  =\Omega\left(  \left(  \frac{1}{1-\mu\left(  g\right)  }\right)
^{h/2}\right)  ,
\]
where $\operatorname*{Q}_{2}\left(  f\right)  $\ is the bounded-error quantum
query complexity of $f$ as defined by Beals et al.\ \cite{bbcmw}. \ In other
words, any $\operatorname*{RFS}$ problem with $\mu$\ bounded away from $0$
requires a number of queries exponential in the tree height $h$.

However, there is an essential further part of the argument, which restricts
the values of $\mu\left(  g\right)  $\ itself. \ Suppose there existed a
family $\left\{  g_{n}\right\}  $\ of `pseudoparity' functions: that is,
$\mu\left(  g_{n}\right)  >0$ for all $n$,\ yet $\mu\left(  g_{n}\right)
=O(1/\log n)$. \ Then the best bound obtainable from Theorem \ref{rfslb}%
\ would be $\Omega\left(  \left(  1+1/\log n\right)  ^{h/2}\right)  $,
suggesting that $\operatorname*{RFS}_{\log^{2}n}^{g}$ might still be solvable
in quantum polynomial time. \ On the other hand, it would be unclear a priori
how to solve $\operatorname*{RFS}_{\log^{2}n}^{g}$\ classically with a
logarithmic number of alternations. \ Theorem \ref{egood} will rule out this
scenario by showing that pseudoparity\ functions do not exist: if $\mu\left(
g\right)  <0.146$ then $g$ is a parity function, and hence $\mu\left(
g\right)  =0$.

The theorem of Ambainis\ that we need is his \textquotedblleft most
general\textquotedblright\ lower bound from \cite{ambainis}, which he
introduced to show that the quantum query complexity of inverting a
permutation is $\Omega\left(  \sqrt{n}\right)  $. \ That theorem can be stated
as follows.

\begin{theorem}
[Ambainis]\label{ambthmrfs}Let $X\subseteq f^{-1}\left(  0\right)  $\ and
$Y\subseteq f^{-1}\left(  1\right)  $\ be sets of inputs to function $f$.
\ Let $R\left(  x,y\right)  \geq0$ be a symmetric real-valued relation
function, and for $x\in X$, $y\in Y$, and index $i$, let%
\begin{align*}
\theta\left(  x,i\right)   &  =\frac{\sum_{y^{\ast}\in Y~:~x\left[  i\right]
\neq y^{\ast}\left[  i\right]  }R\left(  x,y^{\ast}\right)  }{\sum_{y^{\ast
}\in Y}R\left(  x,y^{\ast}\right)  },\\
\theta\left(  y,i\right)   &  =\frac{\sum_{x^{\ast}\in X~:~x^{\ast}\left[
i\right]  \neq y\left[  i\right]  }R\left(  x^{\ast},y\right)  }{\sum
_{y^{\ast}\in Y}R\left(  x^{\ast},y\right)  },
\end{align*}
where the denominators are all nonzero. \ Then $\operatorname*{Q}_{2}\left(
f\right)  =O\left(  1/\upsilon\right)  $\ where%
\[
\upsilon=\max_{x\in X,~y\in Y,~i~:~R\left(  x,y\right)  >0,~x\left[  i\right]
\neq y\left[  i\right]  }\sqrt{\theta\left(  x,i\right)  \theta\left(
y,i\right)  }.
\]

\end{theorem}

We are now ready to prove a lower bound for $\operatorname*{RFS}$.

\begin{theorem}
\label{rfslb}For all $g$ (partial or total), $\operatorname*{Q}_{2}\left(
\operatorname*{RFS}_{h}^{g}\right)  =\Omega\left(  \left(  1-\mu\left(
g\right)  \right)  ^{-h/2}\right)  $.
\end{theorem}

\begin{proof}
Let $X$ be the set of all $0$-inputs to $\operatorname*{RFS}_{h}^{g}$, and let
$Y$ be the set of all $1$-inputs. \ We will weight the inputs using the
distributions $D_{0},D_{1}$\ from the definition of the nonparity coefficient
$\mu\left(  g\right)  $. \ For all $x\in X$, let $p\left(  x\right)  $\ be the
product, over all vertices $v$ in the $\operatorname*{RFS}$\ tree for $x$, of
the probability of the secret string $s$ at $v$, if $s$ is drawn from
$D_{g\left(  s\right)  }$ (where we condition on $v$'s output bit, $g\left(
s\right)  $). \ Next, say that $x\in X$\ and $y\in Y$ \textit{differ
minimally} if, for all vertices $v$ of the $\operatorname*{RFS}$ tree, the
subtrees rooted at $v$ are identical in $x$ and in $y$\ whenever the answer
bit $g\left(  s\right)  $\ at $v$ is the same in $x$ and in $y$.\ \ If $x$ and
$y$ differ minimally, then we will set $R\left(  x,y\right)  =p\left(
x\right)  p\left(  y\right)  $; otherwise we will set $R\left(  x,y\right)
=0$. \ Clearly $R\left(  x,y\right)  =R\left(  y,x\right)  $\ for all $x\in
X,y\in Y$. \ Furthermore, we claim that $\theta\left(  x,i\right)
\theta\left(  y,i\right)  \leq\left(  1-\mu\left(  g\right)  \right)  ^{h}%
$\ for all $x,y$\ that differ minimally and all $i$ such that $x\left[
i\right]  \neq y\left[  i\right]  $. \ For suppose $y^{\ast}\in Y$\ is chosen
with probability proportional to $R\left(  x,y^{\ast}\right)  $, and $x^{\ast
}\in X$\ is chosen with probability proportional to $R\left(  x^{\ast
},y\right)  $. \ Then $\theta\left(  x,i\right)  \theta\left(  y,i\right)
$\ equals the probability that we would notice the switch from $x$\ to
$y^{\ast}$\ by monitoring $i$, times the probability that we would notice the
switch from $y$\ to $x^{\ast}$.

Let $v_{j}$\ be the $j^{th}$ vertex along the path in the $\operatorname*{RFS}%
$\ tree\ from the root to the leaf vertex $i$, for all $j\in\left\{
1,\ldots,h\right\}  $. \ Also, let $z_{j}\in\left\{  0,1\right\}  ^{n}$\ be
the label of the edge between $v_{j-1}$\ and $v_{j}$, and let $s_{x,j}$\ and
$s_{y,j}$\ be the secret strings at $v_{j}$\ in $x$\ and $y$\ respectively.
\ Then since $x$ and $y$ differ minimally, we must have $g\left(
s_{x,j}\right)  \neq g\left(  s_{y,j}\right)  $\ for all $j$---for otherwise
the subtrees rooted at $v_{j}$\ would be identical, which contradicts the
assumption $x\left[  i\right]  \neq y\left[  i\right]  $. \ So we can think of
the process of choosing $y^{\ast}$ as first choosing a random $s_{x,1}%
^{\prime}$\ from $D_{1}$ so that $1=g\left(  s_{x,1}^{\prime}\right)  \neq
g\left(  s_{x,1}\right)  =0$, then choosing a random $s_{x,2}^{\prime}$\ from
$D_{1-g\left(  s_{x,2}\right)  }$ so that $g\left(  s_{x,2}^{\prime}\right)
\neq g\left(  s_{x,2}\right)  $, and so on. \ Choosing $x^{\ast}$ is
analogous, except that whenever we used $D_{0}$\ in choosing $y^{\ast}$\ we
use $D_{1}$, and vice versa. \ Since the $2h$\ secret strings $s_{x,1}%
,\ldots,s_{x,h},s_{y,1},\ldots,s_{y,h}$\ to be updated are independent of one
another, it follows that%
\begin{align*}
\Pr\left[  y^{\ast}\left[  i\right]  \neq x\left[  i\right]  \right]
\Pr\left[  x^{\ast}\left[  i\right]  \neq y\left[  i\right]  \right]   &  =%
{\displaystyle\prod\limits_{j=1}^{h}}
\Pr_{s\in D_{0}}\left[  s\cdot z_{j}\not \equiv s_{x,j}\cdot z_{j}\right]
\Pr_{s\in D_{1}}\left[  s\cdot z_{j}\not \equiv s_{y,j}\cdot z_{j}\right] \\
&  \leq%
{\displaystyle\prod\limits_{j=1}^{h}}
\left(  1-\mu\left(  g\right)  \right) \\
&  =\left(  1-\mu\left(  g\right)  \right)  ^{h}%
\end{align*}
by the definition of $\mu\left(  g\right)  $. \ Therefore%
\[
\operatorname*{Q}\nolimits_{2}\left(  \operatorname*{RFS}\nolimits_{h}%
^{g}\right)  =\Omega\left(  \left(  1-\mu\left(  g\right)  \right)
^{-h/2}\right)
\]
by Theorem \ref{ambthmrfs}.
\end{proof}

Before continuing further, let me show that there is a natural, explicit
choice of $g$---the function $g_{\operatorname{mod}3}\left(  s\right)  $\ from
Section \ref{PRELIMRFS}---for which the nonparity coefficient is almost $3/4$.
\ Thus, for $g=g_{\operatorname{mod}3}$, the algorithm of Lemma \ref{alg}\ is
essentially optimal.

\begin{proposition}
\label{gns}$\mu\left(  g_{\operatorname{mod}3}\right)  =3/4-O\left(
1/n\right)  $.
\end{proposition}

\begin{proof}
Let $n\geq6$. \ Let $D_{0}$ be the uniform distribution over all $s$ with
$\left\vert s\right\vert =3\left\lfloor n/6\right\rfloor $ (so
$g_{\operatorname{mod}3}\left(  s\right)  =0$); likewise let $D_{1}$\ be the
uniform distribution over $s$ with $\left\vert s\right\vert =3\left\lfloor
n/6\right\rfloor +2$ ($g_{\operatorname{mod}3}\left(  s\right)  =1$). \ We
consider only the case of $s$ drawn from $D_{0}$; the $D_{1}$\ case is
analogous. \ We will show that for any $z$,
\[
\left\vert \Pr_{s\in D_{0}}\left[  s\cdot z\equiv0\right]  -\frac{1}%
{2}\right\vert =O\left(  \frac{1}{n}\right)
\]
(all congruences are $\operatorname{mod}2$). \ The theorem then follows, since
by the definition of the nonparity coefficient, given any $z$ the choices of
$s_{0}\in D_{0}$\ and $s_{1}\in D_{1}$\ are independent.

Assume without loss of generality that $1\leq\left\vert z\right\vert \leq n/2$
(if $\left\vert z\right\vert >n/2$, then replace $z$ by its complement). \ We
apply induction on $\left\vert z\right\vert $. \ If $\left\vert z\right\vert
=1$, then clearly%
\[
\Pr\left[  s\cdot z\equiv0\right]  =3\left\lfloor n/6\right\rfloor /n=\frac
{1}{2}\pm O\left(  \frac{1}{n}\right)  \text{.}%
\]
For $\left\vert z\right\vert \geq2$,\ let $z=z_{1}\oplus z_{2}$, where $z_{2}%
$\ contains only the rightmost $1$ of $z$ and $z_{1}$ contains all the other
$1$'s. \ Suppose the proposition holds for $\left\vert z\right\vert -1$.
\ Then%
\begin{align*}
\Pr\left[  s\cdot z\equiv0\right]  =  &  \Pr\left[  s\cdot z_{1}%
\equiv0\right]  \Pr\left[  s\cdot z_{2}\equiv0|s\cdot z_{1}\equiv0\right]  +\\
&  \Pr\left[  s\cdot z_{1}\equiv1\right]  \Pr\left[  s\cdot z_{2}%
\equiv1|s\cdot z_{1}\equiv1\right]  \text{,}%
\end{align*}
where%
\[
\Pr\left[  s\cdot z_{1}\equiv0\right]  =\frac{1}{2}+\alpha,\,\,\,\Pr\left[
s\cdot z_{1}\equiv1\right]  =\frac{1}{2}-\alpha
\]
for some $\left\vert \alpha\right\vert =O\left(  1/n\right)  $. \ Furthermore,
even conditioned on $s\cdot z_{1}$, the expected number of $1$'s in $s$
outside of $z_{1}$\ is $\left(  n-\left\vert z_{1}\right\vert \right)  /2\pm
O\left(  1\right)  $ and they are uniformly distributed. \ Therefore%
\[
\Pr\left[  s\cdot z_{2}\equiv b|s\cdot z_{1}\equiv b\right]  =\frac{1}%
{2}+\beta_{b}%
\]
for some $\left\vert \beta_{0}\right\vert ,\left\vert \beta_{1}\right\vert
=O\left(  1/n\right)  $. \ So%
\begin{align*}
\Pr\left[  s\cdot z\equiv0\right]   &  =\frac{1}{2}+\frac{\beta_{0}}{2}%
+\alpha\beta_{0}-\frac{\beta_{1}}{2}-\alpha\beta_{1}\\
&  =\frac{1}{2}\pm O\left(  \frac{1}{n}\right)  .
\end{align*}

\end{proof}

Finally it must be shown that pseudoparity functions do not exist.\ \ That is,
if $g$ is too close to a parity function for the bound of Theorem
\ref{rfslb}\ to apply, then $g$ actually \textit{is} a parity function, from
which it follows that $RFS_{h}^{g}$\ admits an efficient classical algorithm.

\begin{theorem}
\label{egood}Suppose $\mu\left(  g\right)  <0.146$. \ Then $g$ is a parity
function (equivalently, $\mu\left(  g\right)  =0$).
\end{theorem}

\begin{proof}
By linear programming duality, there exists a joint distribution $\mathcal{D}$
over $z\in\left\{  0,1\right\}  ^{n}$, $0$-inputs $\widehat{s}_{0}\in
g^{-1}\left(  0\right)  $, and $1$-inputs $\widehat{s}_{1}\in g^{-1}\left(
1\right)  $, such that for all $s_{0}\in g^{-1}\left(  0\right)  $ and
$s_{1}\in g^{-1}\left(  1\right)  $,%
\[
\Pr_{\left(  z,\widehat{s}_{0},\widehat{s}_{1}\right)  \in\mathcal{D}}\left[
s_{0}\cdot z\equiv\widehat{s}_{1}\cdot z\left(  \operatorname{mod}2\right)
\,\,\,\vee\,\,\,s_{1}\cdot z\equiv\widehat{s}_{0}\cdot z\left(
\operatorname{mod}2\right)  \right]  <\mu\left(  g\right)  \text{.}%
\]
Furthermore $\widehat{s}_{0}\cdot z\not \equiv \widehat{s}_{1}\cdot z\left(
\operatorname{mod}2\right)  $, since otherwise we could violate the hypothesis
by taking $s_{0}=\widehat{s}_{0}$\ or $s_{1}=\widehat{s}_{1}$. \ It follows
that there exists a joint distribution $\mathcal{D}^{\prime}$\ over
$z\in\left\{  0,1\right\}  ^{n}$ and $b\in\left\{  0,1\right\}  $ such that%
\[
\Pr_{\left(  z,b\right)  \in\mathcal{D}^{\prime}}\left[  s\cdot z\equiv
b\left(  \operatorname{mod}2\right)  \right]  >1-\mu\left(  g\right)
\]
for all $s\in g^{-1}\left(  0\right)  $, and%
\[
\Pr_{\left(  z,b\right)  \in\mathcal{D}^{\prime}}\left[  s\cdot z\not \equiv
b\left(  \operatorname{mod}2\right)  \right]  >1-\mu\left(  g\right)
\]
for all $s\in g^{-1}\left(  1\right)  $. \ But this implies that $g$ is a
bounded-error threshold function of parity functions. \ More precisely, there
exist probabilities $p_{z}$, summing to $1$, as well as $b_{z}\in\left\{
0,1\right\}  $\ such that for all $s\in\left\{  0,1\right\}  ^{n}$,%
\[
\Psi\left(  s\right)  =\sum_{z\in\left\{  0,1\right\}  ^{n}}p_{z}\left(
\left(  s\cdot z\right)  \oplus b_{z}\right)  \text{ is }\left\{
\begin{array}
[c]{ll}%
>1-\mu\left(  g\right)  & \text{if }g\left(  s\right)  =1\\
<\mu\left(  g\right)  & \text{if }g\left(  s\right)  =0.
\end{array}
\ \ \right.
\]
We will consider $\operatorname*{var}\left(  \Psi\right)  $, the variance of
the above quantity $\Psi\left(  s\right)  $\ if $s$ is drawn uniformly at
random from $\left\{  0,1\right\}  ^{n}$. \ First, if $p_{z}\geq1/2$\ for any
$z$, then $g\left(  s\right)  =\left(  s\cdot z\right)  \oplus b_{z}$\ is a
parity function and hence $\mu\left(  g\right)  =0$. \ So we can assume
without loss of generality that $p_{z}<1/2$\ for all $z$. \ Then since $s$ is
uniform, for each $z_{1}\neq z_{2}$\ we know that $\left(  s\cdot
z_{1}\right)  \oplus b_{z_{1}}$\ and $\left(  s\cdot z_{2}\right)  \oplus
b_{z_{2}}$\ are pairwise independent $\left\{  0,1\right\}  $ random
variables, both with expectation $1/2$. \ So%
\[
\operatorname*{var}\left(  \Psi\right)  =\frac{1}{4}%
{\textstyle\sum\nolimits_{z}}
p_{z}^{2}<\frac{1}{4}\left(  \left(  \frac{1}{2}\right)  ^{2}+\left(  \frac
{1}{2}\right)  ^{2}\right)  =\frac{1}{8}\text{.}%
\]
On the other hand, since $\Psi\left(  s\right)  $\ is always less than $\mu
$\ or greater than $1-\mu$,%
\[
\operatorname*{var}\left(  \Psi\right)  >\left(  \frac{1}{2}-\mu\right)
^{2}.
\]
Combining,%
\[
\mu>\frac{2-\sqrt{2}}{4}>0.146.
\]

\end{proof}

\section{Open Problems\label{OPENRFS}}

An intriguing open problem is whether Theorem \ref{rfslb}\ can be proved using
the polynomial method of Beals et al.\ \cite{bbcmw}, rather than the adversary
method of Ambainis \cite{ambainis}. \ It is known that one can lower-bound
polynomial degree in terms of block sensitivity, or the maximum number of
disjoint changes to an input that change the output value. \ The trouble is
that the $\operatorname*{RFS}$ function has block sensitivity $1$---the
\textquotedblleft sensitive blocks\textquotedblright\ of each input tend to
have small intersection, but are not disjoint. \ For this reason, I implicitly
used the \textquotedblleft quantum certificate complexity\textquotedblright%
\ as defined in \cite{aar:cer} rather than block sensitivity to prove a lower bound.

I believe the constant of Theorem \ref{egood}\ can be improved. \ The smallest
nonzero $\mu\left(  g\right)  $\ value I know of is attained when $n=2$ and
$g=\operatorname*{OR}\left(  s\left[  1\right]  ,s\left[  2\right]  \right)  $:

\begin{proposition}
$\mu\left(  \operatorname*{OR}\right)  =1/3$.
\end{proposition}

\begin{proof}
First, $\mu\left(  \operatorname*{OR}\right)  \geq1/3$, since $D_{1}$\ can
choose $s\left[  1\right]  s\left[  2\right]  $\ to be $01$, $10$,\ or $11$
each with probability $1/3$; then for any $z\neq0$ and the unique $0$-input
$\widehat{s}_{0}=00$, we have $s_{1}\cdot z\not \equiv \widehat{s}_{0}\cdot z$
with probability at most $2/3$. \ Second, $\mu\left(  \operatorname*{OR}%
\right)  \leq1/3$, since applying linear programming duality, we can let the
pair $\left(  z,\widehat{s}_{1}\right)  $ equal $\left(  01,01\right)  $,
$\left(  10,10\right)  $, or $\left(  11,10\right)  $\ each with probability
$1/3$. \ Then $0\equiv s_{0}\cdot z\not \equiv \widehat{s}_{1}\cdot z\equiv
1$\ always, and for any $1$-input $s_{1}$, we have $s_{1}\cdot z\equiv
1\not \equiv \widehat{s}_{0}\cdot z$ with probability $2/3$.
\end{proof}

Finally, I conjecture that uncomputation is unavoidable not just for
$\operatorname*{RFS}$ but for many other recursive problems, such as game-tree
evaluation. \ Formally, the conjecture is that the quantum query complexity of
evaluating a game tree increases exponentially with depth as the number of
leaves is held constant, even if there is at most one winning move per vertex
(so that the tree can be evaluated with zero probability of error).

\section{Acknowledgments}

I thank Lisa Hales, Umesh Vazirani, Ronald de Wolf, and the
anonymous reviewers for helpful comments. \ This work was done while
I was a graduate student at UC Berkeley, supported by an NSF
Graduate Fellowship.

\bibliographystyle{plain}
\bibliography{thesis}

\end{document}